\documentclass[a4paper,10pt,twoside]{cpc-hepnp}

\usepackage{multicol}
\usepackage{graphicx}
\usepackage{booktabs}
\usepackage{amssymb,bm,mathrsfs,bbm,amscd}
\usepackage[tbtags]{amsmath}
\usepackage{lastpage}

\usepackage{ifpdf}
\usepackage{feynmp}
\ifpdf
\else
\fi

\begin{document}

\fancyhead[c]{\small Chinese Physics C~~~Vol. xx, No. x (201x) xxxxxx}
\fancyfoot[C]{\small 010201-\thepage}


\title{Physics cross sections and event generation of $e^+e^-$ annihilations at the CEPC
\thanks{The study was partially supported by the CAS/SAFEA International Partnership Program for Creative Research Teamsㄛand funding from CAS and IHEP for the Thousand Talent and Hundred Talent programs, as well as grants from the State Key Laboratory of Nuclear Electronics and Particle Detectors.}}

\author{%
      Xin. Mo$^{1;1)}$\email{moxin@ihep.ac.cn}%
\quad Gang. Li$^{1;2)}$\email{li.gang@ihep.ac.cn}%
\quad Manqi. Ruan$^{1)}$%
\quad Xinchou. Lou$^{1,2)}$%
}
\maketitle

\address{%
$^1$ Institute of High Energy Physics, Chinese Academy of Sciences, Beijing 100049, China\\
$^2$ University of Texas at Dallas, Richardson, TX 75080-3021, USA
}

\begin{abstract}
The cross sections of the Higgs production and the corresponding backgrounds of $e^+e^-$ annihilations at the CEPC (Circular Electron and Positron Collider) are calculated by a Monte-Carlo method, and the beamstruhlung effect at the CEPC is carefully investigated. The potential for new physics beyond the Standard Model at the CEPC is discussed.
\end{abstract}



\footnotetext[0]{\hspace*{-3mm}\raisebox{0.3ex}{$\scriptstyle\copyright$}2013
Chinese Physical Society and the Institute of High Energy Physics
of the Chinese Academy of Sciences and the Institute
of Modern Physics of the Chinese Academy of Sciences and IOP Publishing Ltd}%

\begin{multicols}{2}

\section{Introduction}
The discovery of the Higgs boson~\cite{ref:1,ref:2} is a great milestone for the modern particle physics. Whether it is the Standard Model (SM) Higgs or an indication of new physics, the fact  that it is found in the search of the SM Higgs boson  makes the Higgs mechanism  more creditable. Furthermore, precision measurements of properties of the new Higgs boson are critical for the Higgs physics; any deviation away from the SM expectation will improve our knowledge of the elementary particles and their interactions. With this consideration, an $e^+e^-$ collider with high luminosity and energy is best suited for the Higgs research.

The Circular Electron-Positron Collider(CEPC)~\cite{ref:3} is a proposed circular collider. It is designed to run around $240\sim250$ GeV and will deliver 5 $\mathrm{ab}^{-1}$ of integrated luminosity during 10 years of operation. About $10^6$ Higgs events will be produced in an clean environment,  which allows the measurement of the cross section of the Higgs production as well as its mass, decay width and branching ratios with precision much beyond those of hadron colliders.

Besides the Higgs events, there will be a large number of electroweak processes generated at the CEPC, via the vector bosons W and Z. The large electroweak statistics is essential for understanding the detectors precisely, for the precision measurements of Higgs  properties. Moreover,  these electroweak events can also be used to perform high precision measurements of the SM parameters, such as  the forward-backward charge asymmetry ($A_{FB}$).


The calculation in this paper is done with the {\sc Whizard} generator~\cite{ref:4,ref:5},  which is one of the most popular Monte-Carlo generators for the Higgs study and  is employed by ILC and CLIC groups.

In this paper, we will investigate the physics cross sections and describe the event generation at the CEPC. In Sec. \uppercase\expandafter{\romannumeral2} the Higgs production is studied and the cross sections are given. The radiative correction to the cross sections is discussed in Sec. \uppercase\expandafter{\romannumeral3}. In Sec. \uppercase\expandafter{\romannumeral4}, the backgrounds of various final states are presented. In the end, a brief summary is provided in the Sec. \uppercase\expandafter{\romannumeral5}.

\section{Higgs production cross section}
\begin{center}
\vspace{0.2cm}
\begin{fmffile}{zh_i1_diags-fmf}
\fmfframe(8,7)(8,6){%
  \begin{fmfgraph*}(70,60)
   \fmfpen{thin}
   \fmfset{arrow_len}{2mm}
    \fmfleft{i1,i2}
    \fmfright{o3,o4}
    \fmflabel{${e^-}$}{i1}
    \fmflabel{${e^+}$}{i2}
    \fmflabel{${Z}$}{o3}
    \fmflabel{${H}$}{o4}
    \fmf{fermion}{i1,v1}
    \fmfdot{v1}
    \fmfdot{v1}
    \fmf{fermion}{v1,i2}
    \fmf{wiggly,label=\begin{scriptsize}${Z^*}$\end{scriptsize}}{v1,v5}
    \fmfdot{v1,v5}
    \fmfdot{v5}
    \fmf{dashes,tension=0.5}{v5,o4}
    \fmfdot{v5}
    \fmf{wiggly,tension=0.5}{v5,o3}
  \end{fmfgraph*}}\hfil\allowbreak
\end{fmffile}
\vspace{0.6cm}
\figcaption{\label{fig:zh}The Feynman diagram of the Higgsstrahlung process}
\end{center}

Due to the fact that the Higgs boson coupling to fermions is proportional to the fermion mass,
the process $e^+e^-\to H$ is highly suppressed.
The dominant Higgs production in electron positron annihilation is the so-called Higgsstrahlung,
a $s$-channel process in which Higgs is produced in association with a $Z$ boson, as shown in Fig.~\ref{fig:zh}.

The other two typical $t$-channel processes, the vector boson fusions, are sizable contributions to the Higgs production that will increase with the center of mass energy.
\begin{center}
\vspace{0.2cm}
\begin{fmffile}{wf_i1_diags-fmf}
\fmfframe(8,7)(8,6){%
  \begin{fmfgraph*}(70,60)
   \fmfpen{thin}
   \fmfset{arrow_len}{2mm}
    \fmfleft{i1,i2}
    \fmfright{o3,o5,o4}
    \fmflabel{${e^-}$}{i1}
    \fmflabel{${e^+}$}{i2}
    \fmflabel{${\nu_e}$}{o3}
    \fmflabel{${H}$}{o5}
    \fmflabel{${\bar{\nu}_e}$}{o4}
    \fmf{fermion}{i1,v1}
    \fmfdot{v1}
    \fmf{wiggly,label=\begin{scriptsize}${W}$\end{scriptsize}}{v1,v4}
    \fmfdot{v1,v4}
    \fmf{wiggly,label=\begin{scriptsize}${W}$\end{scriptsize}}{v4,v16}
    \fmfdot{v4,v16}
    \fmfdot{v16}
    \fmf{fermion}{v16,i2}
    \fmfdot{v16}
    \fmf{fermion,tension=0.5}{o4,v16}
    \fmfdot{v4}
    \fmf{dashes,tension=0.5}{v4,o5}
    \fmfdot{v1}
    \fmf{fermion,tension=0.5}{v1,o3}
  \end{fmfgraph*}}\hfil\allowbreak
\end{fmffile}
\begin{fmffile}{zf_i1_diags-fmf}
\fmfframe(8,7)(8,6){%
  \begin{fmfgraph*}(70,60)
   \fmfpen{thin}
   \fmfset{arrow_len}{2mm}
    \fmfleft{i1,i2}
    \fmfright{o3,o5,o4}
    \fmflabel{${e^-}$}{i1}
    \fmflabel{${e^+}$}{i2}
    \fmflabel{${e^-}$}{o3}
    \fmflabel{${H}$}{o5}
    \fmflabel{${e^+}$}{o4}
    \fmf{fermion}{i1,v1}
    \fmfdot{v1}
    \fmf{wiggly,label=\begin{scriptsize}${Z}$\end{scriptsize}}{v1,v4}
    \fmfdot{v1,v4}
    \fmf{wiggly,label=\begin{scriptsize}${Z}$\end{scriptsize}}{v4,v16}
    \fmfdot{v4,v16}
    \fmfdot{v16}
    \fmf{fermion}{v16,i2}
    \fmfdot{v16}
    \fmf{fermion,tension=0.5}{o4,v16}
    \fmfdot{v4}
    \fmf{dashes,tension=0.5}{v4,o5}
    \fmfdot{v1}
    \fmf{fermion,tension=0.5}{v1,o3}
  \end{fmfgraph*}}\hfil\allowbreak
\end{fmffile}
\vspace{0.2cm}
\figcaption{\label{fig:vbf}The Feynman diagrams of vector boson fusions}
\end{center}

The cross section of the Higgsstrahlung can be written as
\begin{equation}\label{eq:1}
\sigma(e^+e^-\to ZH)=\frac{G^2_Fm^4_Z}{96\pi s}(v^2_e+a^2_e)\lambda^{\frac{1}{2}}\frac{\lambda+12m^2_Z/s}{(1-m^2_Z/s)^2}.
\end{equation}
And the cross section for the vector fusion production can be written in a similar compact form as~\cite{ref:6}
\begin{equation}\label{eq:2}
\sigma(e^+e^-\to VV\to l\bar{l}H)=\frac{G^3_Fm^4_V}{4\sqrt{2}\pi^3}\int_{x_H}^1\mathrm{d}x\int_x^1\frac{\mathrm{d}y F(x,y)}{\left[1+(y-x)/x_V^2\right]^2},\\
\end{equation}
\begin{eqnarray*}
F(x,y)&=&\left(\frac{2x}{y^3}-\frac{3x+1}{y^2}+\frac{x+2}{y}-1\right)\left[\frac{z}{z+1}-\mathrm{log}(z+1)\right]\\
&&+\frac{x z^2(1-y)}{y^3(z+1)},
\end{eqnarray*}
where V stands for vector bosons $Z$ or $W$, and the dimensionless variables are defined as
$$x_H=m^2_H/s, \quad x_V=m^2_V/s \quad z=y(x-x_H)/(xx_V).$$

For the moderate Higgs masses and energies, the cross sections of vector boson fusion are suppressed compared with the Higgsstrahlung due to the additional electroweak couplings. With the increasing energy, the cross sections of VBF processes rise logarithmically as the typical $t$-channel process~\cite{ref:7}

\begin{equation}\label{eq:2}
\sigma(e^+e^-\to VV\to l\bar{l}H)\approx\frac{G^3_Fm^4_V}{4\sqrt{2}\pi^3}\mathrm{log}\frac{s}{m^2_H}.\\
\end{equation}
Meanwhile, the cross section of the Higgsstrahlung decreases according to the scaling law $\sim g^4_F/s$ asymptotically.
Therefore the vector boson fusion will become the dominant contribution to the Higgs production far beyond the $ZH$ threshold.

\begin{center}
\includegraphics[width=0.9\linewidth]{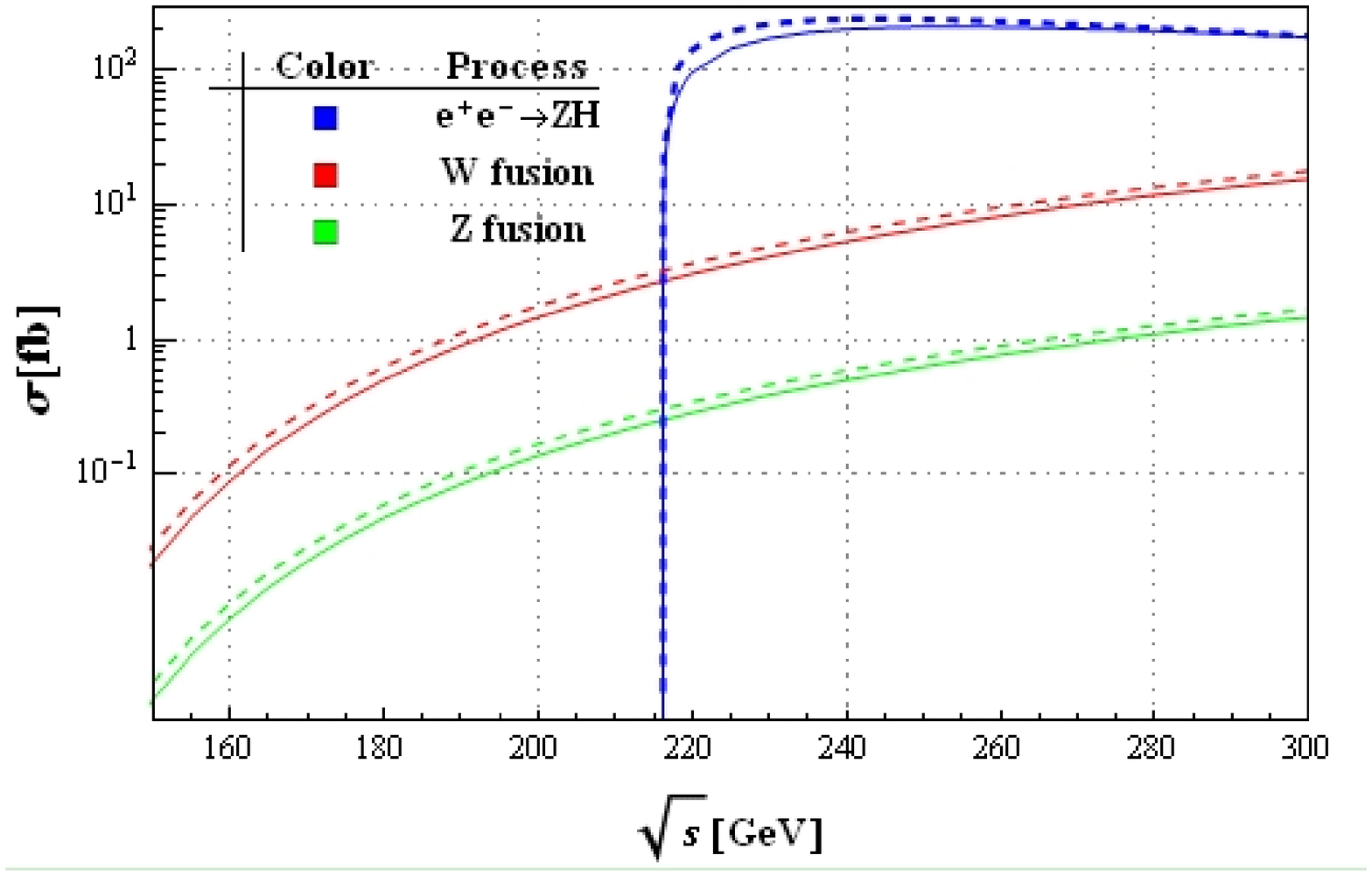}
\figcaption{\label{fig:higgprod} The cross sections of various Higgs productions, in which the solid lines are with ISR correction and the dashed ones are without.}
\end{center}

The Higgsstrahlung, the $WW$ and $ZZ$ fusions are shown in Fig.~\ref{fig:higgprod}. Above the threshold of $ZH$, the cross section of Higgsstrahlung rises rapidly and reaches its maximum around $250 \mathrm{GeV}$. The dashed curves represent the cross sections without ISR effect which will be discussed in the following section.

\section{Initial state radiation and beamstrahlung}

The Initial State Radiation (ISR) is an important issue in high energy processes, especially for lepton colliders.
ISR affects cross section significantly, for example, reduces the $ZH$ cross section by more than 10\% within the {\sc Whizard} framework as shown in Fig.~\ref{fig:higgprod}.
Though the cancelation for soft and collinear photons has been validated to all orders in perturbation theory~\cite{ref:8}, and the hard photon emission could also be calculated perturbatively order by order in QED in principle~\cite{ref:9,ref:10}, actually, the maximal order for the hard photon is set to be 3 in the calculation.

\begin{center}
\tabcaption{Comparison of cross sections with beamstrahlung turned on and off \label{tab:beam}}
\begin{tabular}{@{}*{3}{lcc}}
\toprule
                       &  ISR [fb]   & ISR \& Beamstrahlung [fb] \\ \hline
$\sigma(e^+e^-\to ZH)$  & 212 & 211                     \\
$\sigma(e^+e^-\to \nu\bar{\nu}H)$  & 6.72 & 6.72                     \\
$\sigma(e^+e^-\to e^+e^-H)$  & 0.63 & 0.63                     \\
$\sigma(e^+e^-\to q\bar{q})$  & 50216 & 50416                     \\
$\sigma(e^+e^-\to W^+W^-)$  & 15484 & 15440                     \\
$\sigma(e^+e^-\to ZZ)$  & 1033 & 1030                     \\
\toprule
\end{tabular}
\end{center}

Besides ISR, another macro effect at high luminosity electron-positron collider, beamstrahlung,   also affects the cross section.  In the storage ring the beamstrahlung  effect makes the beam energy spread larger and reduces the center of mass energy\cite{ref:cepc_acc}.
The tool that is widely used to simulate this effect for $e^+e^-$ colliders is GuineaPig++~\cite{ref:11,ref:12}.
The total energy spread caused by beamstrahlung and synchrotron radiation is studied by Monte-Carlo simulation and determined to be 0.1629\% at CEPC~\cite{ref:cepc_acc}.  In the event generation, even though the energy spread is taken into account,  the correction on the cross sections of Higgs production is negligible while the observable difference on the $e^{+}e^{-}\to q\bar{q}$ is only at percentage level. The results are  listed in Tab.~\ref{tab:beam}.


\section{Cross sections of backgrounds}
\subsection{Two fermions backgrounds}

Two-fermion final states are the major backgrounds in electron positron collider. The leading process is the quark pair production and the cross section is around 50~pb at $\sqrt{s}\sim250$GeV. Another notable two-fermion process is the di-muon production, whose cross section reaches 5~pb. The expected numbers of $q\bar{q}$ and $\mu^{+}\mu^{-}$ events at 250 GeV are around $ 250\times 10^{6}$ and $ 22\times 10^{6}$, respectively(see Tab.~\ref{tab:sam}), for the CEPC program.

\begin{center}
\includegraphics[width=0.9\linewidth]{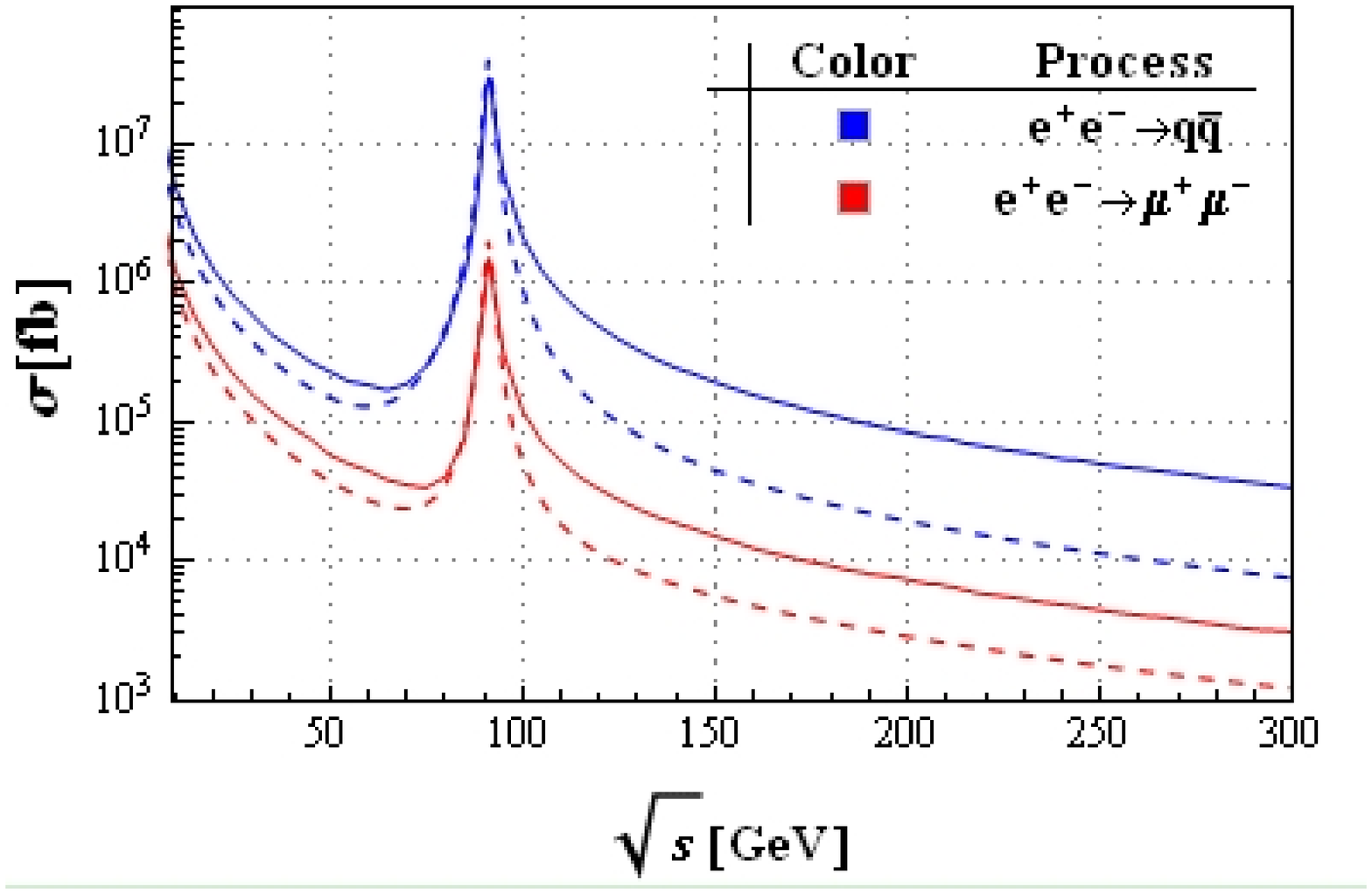}
\figcaption{\label{fig:qqmu} The cross sections of $e^{+}e^{-} \to q\bar{q}$ and $\mu^{+}\mu^{-}$ processes, in which the solid (dashed) lines are with(without) ISR correction.}
\end{center}

The cross sections are much higher in the vicinity of the Z pole and the improvements to the precision of Z pole measurement could benefit from such an large statistics. For the partial decay width $Z\to\mu^+\mu^-$, the statistical uncertainty can be reduced to $0.03\%$~\cite{ref:3} compared to $0.15\%$ for LEP experiments~\cite{ref:13,ref:14,ref:15,ref:16}. The electroweak mixing angle is an important free parameter in SM, which rotates the original $SU(2)\times U(1)$ filed into the physically observed vector boson state. It is sensitive to higher-order electroweak correction and could be a precise test of SM perturbative theory. The precision of the mixing angle is expected to be $0.02\%$ from $Z\to b\bar{b}$ process Ref.~\cite{ref:3}. The precision measurement for the mass of the Z boson may also be improved on CEPC, and the uncertainty due to statistic is expected to be 0.1MeV~\cite{ref:3}.

Both the measurements for electroweak mixing angle and mass of Z boson need the data of Z pole run as well as off-peak runs.
A preliminary data taking plan for the Z pole physics was proposed in ~\cite{ref:3}, the  luminosities as well as the cross sections at different energy points are  listed in Tab.~\ref{tab:run}. Of course, this scan scheme would be optimized in the future study.

\begin{center}
\tabcaption{A possible scan scheme around Z pole on CEPC\label{tab:run}}
\begin{tabular}{@{\extracolsep{\fill}}lcccc}
\toprule
$\sqrt{s}\left[\mathrm{GeV}\right]$ & 88.2 & 89.2 & 90.2 & 91.1876   \\
$\int L\left[\mathrm{fb^{-1}}\right]$    &  10  &  10  &  10  &   100      \\
$\sigma(e^+e^-\to q\bar{q})[\mathrm{nb}]$        & 4.22 & 7.92 & 17.22 & 30.20 \\
$\sigma(e^+e^-\to \mu\mu)[\mathrm{nb}]$          & 0.22 & 0.41 & 0.87  & 1.51  \\
\hline
$\sqrt{s}\left[\mathrm{GeV}\right]$ & 92.2 & 93.2 & 94.2 &         \\
$\int L\left[\mathrm{fb^{-1}}\right]$    &  10  &  10  &  10  &         \\
$\sigma(e^+e^-\to q\bar{q})[\mathrm{nb}]$  & 22.24 & 12.78 & 8.18  \\
$\sigma(e^+e^-\to \mu\mu)[\mathrm{nb}]$    &  1.11 &  0.65 & 0.42  \\
\toprule
\end{tabular}
\end{center}

\subsection{Bhabha scattering}
The small-angle(SABH) and large-angle(LABH) Bhabha scattering processes both play important roles in $e^{+}e^{-}$ collider physics.
SABH is mainly used to measure the instant luminosity of the accelerator due to the huge cross section,
while LABH can be used as a cross check of the luminosity measurement of the SABH  and the calibration of detector during offline stage.

\begin{center}
\includegraphics[width=0.8\linewidth]{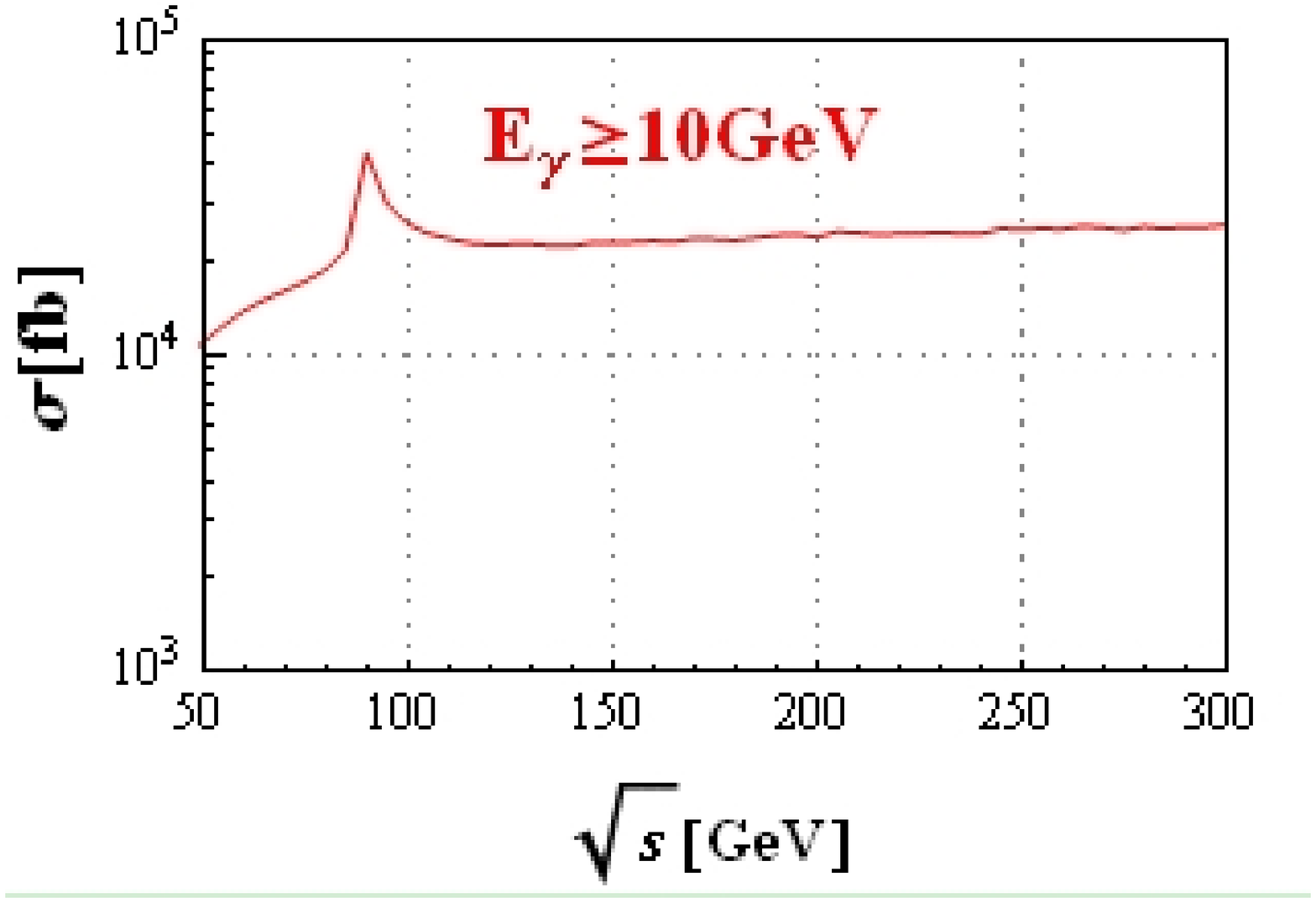}
\figcaption{\label{fig:bha} The cross sections of Bhabha process, ISR is included and a cut for photon energy is applied.}
\end{center}

The observed cross section of SABH in the next-to-leading order for theoretical prediction depends not only on the angular range covered by detector, but also on the cut of the electron energy. Although the {\sc Whizard} is designed to calculate it up to next leading order, unfortunately, it is not adequate for SABH process, especially at the low energy region.



\subsection{Vector boson pair production}
The electroweak processes for vector boson pair production are $e^+e^-\to W^+W^-$ and $e^+e^-\to ZZ$. The cross sections of on-shell production for vector bosons are plotted in Fig.~\ref{fig:wz}. The $s$-channel process provides the dominant contribution in the CEPC energy region.

The vector boson pair production is crucial for the precision measurement of the Higgs.
For example, W bosons can decay into a quark-antiquark pair, i.e., $W^+W^-\to u\bar{u}d\bar{d}\quad \mathrm{or}\quad c\bar{c}s\bar{s}$,
which creates four jets and tends to contaminate the $ZH$ in many cases.
Although ZZ pair production is one magnitude smaller than WW, it is especially severe for the background of the Higgs measurements
as it may lead to the same final states  and  kinematics with $ZH$.

\begin{center}
\includegraphics[width=0.9\linewidth]{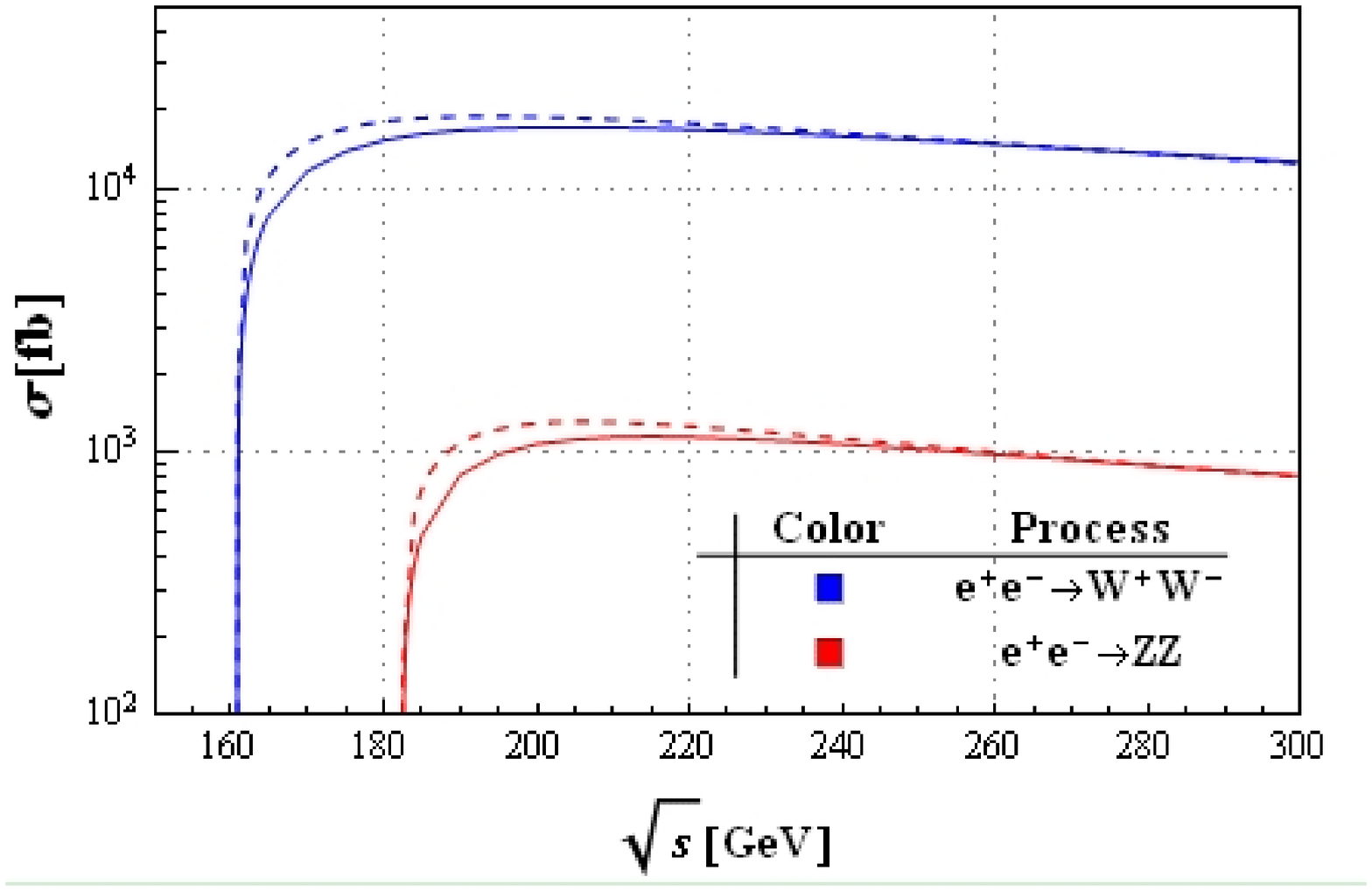}
\figcaption{\label{fig:wz} The cross sections of  $e^{+}e^{-} \to W^{+}W^{-}$ and $ZZ$ processes, in which the solid lines are with ISR correction and the dashed ones without.}
\end{center}

The precision measurements for W boson mass can be achieved by both the direct method and threshold scan.
By applying the direct method, the measurements can be performed with the same run at $\sqrt{s}=250\mathrm{GeV}$ where most attention is paid on Higgs.
The main challenge of direct method is the calibration of  detectors and the modeling of resolution, which is proportional to $\sqrt{s}$. Thanks to the high luminosity of CEPC, the uncertainty due to the beamstrahlung can be reduced to $\sim1\mathrm{MeV}$.


\begin{center}
\begin{fmffile}{ww_i1_diags-fmf}
\fmfframe(8,7)(8,6){%
  \begin{fmfgraph*}(70,60)
   \fmfpen{thin}
   \fmfset{arrow_len}{2mm}
    \fmfleft{i1,i2}
    \fmfright{o3,o4}
    \fmflabel{${e^-}$}{i1}
    \fmflabel{${e^+}$}{i2}
    \fmflabel{${W^+}$}{o3}
    \fmflabel{${W^-}$}{o4}
    \fmf{fermion}{i1,v1}
    \fmfdot{v1}
    \fmfdot{v1}
    \fmf{fermion}{v1,i2}
    \fmf{wiggly,label=\begin{scriptsize}${Z}$\end{scriptsize}}{v1,v5}
    \fmfdot{v1,v5}
    \fmfdot{v5}
    \fmf{wiggly,tension=0.5}{v5,o4}
    \fmfdot{v5}
    \fmf{wiggly,tension=0.5}{v5,o3}
  \end{fmfgraph*}}\hfil\allowbreak
\fmfframe(8,7)(8,6){%
  \begin{fmfgraph*}(70,60)
   \fmfpen{thin}
   \fmfset{arrow_len}{2mm}
    \fmfleft{i1,i2}
    \fmfright{o4,o3}
    \fmflabel{${e^-}$}{i1}
    \fmflabel{${e^+}$}{i2}
    \fmflabel{${W^-}$}{o4}
    \fmflabel{${W^+}$}{o3}
    \fmf{fermion}{i1,v1}
    \fmfdot{v1}
    \fmf{fermion,label=\begin{scriptsize}${\nu_e}$\end{scriptsize}}{v1,v4}
    \fmfdot{v1,v4}
    \fmfdot{v4}
    \fmf{fermion}{v4,i2}
    \fmfdot{v4}
    \fmf{wiggly,tension=0.5}{v4,o3}
    \fmfdot{v1}
    \fmf{wiggly,tension=0.5}{v1,o4}
  \end{fmfgraph*}}\hfil\allowbreak
\end{fmffile}
\figcaption{\label{fig:ww}The Feynman diagrams of WW pair production}
\end{center}

Besides the W boson mass measurements, these vector boson productions, $W^{+}W^{-}$ and $ZZ$, are also a good resource for the precision measurements for triple boson coupling of electroweak theory. Any deviation from SM will be the hint for new physics beyond the Standard Model.


\section{Classification of event samples}
The Monte-Carlo samples generated for CEPC could be grouped into signal part and background part intuitively.
The signal part includes all the Higgs production processes, both Higgsstrahlung and fusion.
The backgrounds could be classified according to the final states furthermore.
The final states with two fermions and four fermions contribute the most of backgrounds for CEPC, since the center of mass energy of CEPC is not high enough for more particles in final state.

\begin{center}
\tabcaption{Number of Feynman diagrams for WW type processes\label{tab:ww}}
\begin{tabular}{@{}*{3}{lccccc}}
\toprule
                       & $u\bar{d}$ & $c\bar{s}$ & $\bar{e}\nu_e$ & $\bar{\mu}\nu_{\mu}$ & $\bar{\tau}\nu_{\tau}$ \\ \hline
$d\bar{u}$             & \emph{43}  & \textbf{11}  &      20      &  \textbf{10}         &  \textbf{10}           \\
$s\bar{c}$             & \textbf{11}& \emph{44}    &      20      &  \textbf{10}         &  \textbf{10}           \\
$e\bar{\nu}_e$         & 20         & 20           &\emph{56}     &     18               &  18                    \\
$\mu\bar{\nu}_{\mu}$   & \textbf{10}& \textbf{10}  &      18      &     \emph{19}        &  \textbf{9}            \\
$\tau\bar{\nu}_{\tau}$ & \textbf{10}& \textbf{10}  &      18      &     \textbf{9}       &  \emph{20}             \\
\toprule
\end{tabular}
\end{center}

\begin{center}
\begin{fmffile}{ww_four_i1_diags-fmf}
\fmfframe(8,7)(8,6){%
  \begin{fmfgraph*}(70,60)
   \fmfpen{thin}
   \fmfset{arrow_len}{2mm}
    \fmfleft{i1,i2}
    \fmfright{o5,o6,o3,o4}
    \fmflabel{${e^-}$}{i1}
    \fmflabel{${e^+}$}{i2}
    \fmflabel{${\mu^-}$}{o5}
    \fmflabel{${\bar{\nu}_\mu}$}{o6}
    \fmflabel{${u}$}{o3}
    \fmflabel{${\bar{d}}$}{o4}
    \fmf{fermion}{i1,v1}
    \fmfdot{v1}
    \fmfdot{v1}
    \fmf{fermion}{v1,i2}
    \fmf{wiggly,label=\begin{scriptsize}${Z}$\end{scriptsize}}{v1,v5}
    \fmfdot{v1,v5}
    \fmf{wiggly,label=\begin{scriptsize}${W}$\end{scriptsize}}{v5,v20}
    \fmfdot{v5,v20}
    \fmfdot{v20}
    \fmf{fermion,tension=0.5}{o4,v20}
    \fmfdot{v20}
    \fmf{fermion,tension=0.5}{v20,o3}
    \fmf{wiggly,label=\begin{scriptsize}${W}$\end{scriptsize}}{v5,v21}
    \fmfdot{v5,v21}
    \fmfdot{v21}
    \fmf{fermion,tension=0.5}{o6,v21}
    \fmfdot{v21}
    \fmf{fermion,tension=0.5}{v21,o5}
  \end{fmfgraph*}}\hfil\allowbreak
\fmfframe(8,7)(8,6){%
  \begin{fmfgraph*}(70,60)
   \fmfpen{thin}
   \fmfset{arrow_len}{2mm}
    \fmfleft{i1,i2}
    \fmfright{o5,o6,o3,o4}
    \fmflabel{${e^-}$}{i1}
    \fmflabel{${e^+}$}{i2}
    \fmflabel{${\mu^-}$}{o5}
    \fmflabel{${\bar{\nu}_\mu}$}{o6}
    \fmflabel{${u}$}{o3}
    \fmflabel{${\bar{d}}$}{o4}
    \fmf{fermion}{i1,v1}
    \fmfdot{v1}
    \fmf{fermion,label=\begin{scriptsize}${}$\end{scriptsize}}{v1,v4}
    \fmfdot{v1,v4}
    \fmfdot{v4}
    \fmf{fermion}{v4,i2}
    \fmf{wiggly,label=\begin{scriptsize}${W}$\end{scriptsize}}{v4,v17}
    \fmfdot{v4,v17}
    \fmfdot{v17}
    \fmf{fermion,tension=0.5}{o4,v17}
    \fmfdot{v17}
    \fmf{fermion,tension=0.5}{v17,o3}
    \fmf{wiggly,label=\begin{scriptsize}${W}$\end{scriptsize}}{v1,v5}
    \fmfdot{v1,v5}
    \fmfdot{v5}
    \fmf{fermion,tension=0.5}{o6,v5}
    \fmfdot{v5}
    \fmf{fermion,tension=0.5}{v5,o5}
  \end{fmfgraph*}}\hfil\allowbreak
  \vspace{15pt}
\fmfframe(8,7)(8,6){%
  \begin{fmfgraph*}(70,60)
   \fmfpen{thin}
   \fmfset{arrow_len}{2mm}
    \fmfleft{i1,i2}
    \fmfright{o5,o4,o6,o3}
    \fmflabel{${e^-}$}{i1}
    \fmflabel{${e^+}$}{i2}
    \fmflabel{${\mu}$}{o5}
    \fmflabel{${\bar{\nu}_{\mu}}$}{o4}
    \fmflabel{${\bar{d}}$}{o6}
    \fmflabel{${u}$}{o3}
    \fmf{fermion}{i1,v1}
    \fmfdot{v1}
    \fmfdot{v1}
    \fmf{fermion}{v1,i2}
    \fmf{boson,label=\begin{scriptsize}${}$\end{scriptsize}}{v1,v5}
    \fmfdot{v1,v5}
    \fmf{fermion,label=\begin{scriptsize}${}$\end{scriptsize}}{v20,v5}
    \fmfdot{v5,v20}
    \fmf{boson,label=\begin{scriptsize}${W}$\end{scriptsize}}{v20,v80}
    \fmfdot{v20,v80}
    \fmfdot{v80}
    \fmf{fermion,tension=0.5}{v80,o3}
    \fmfdot{v80}
    \fmf{fermion,tension=0.5}{o6,v80}
    \fmfdot{v20}
    \fmf{fermion,tension=0.5}{o4,v20}
    \fmfdot{v5}
    \fmf{fermion,tension=0.5}{v5,o5}
  \end{fmfgraph*}}\hfil\allowbreak
\fmfframe(8,7)(8,6){%
  \begin{fmfgraph*}(70,60)
   \fmfpen{thin}
   \fmfset{arrow_len}{2mm}
    \fmfleft{i1,i2}
    \fmfright{o5,o4,o6,o3}
    \fmflabel{${e^-}$}{i1}
    \fmflabel{${e^+}$}{i2}
    \fmflabel{${u}$}{o5}
    \fmflabel{${\bar{d}}$}{o4}
    \fmflabel{${\bar{\nu}_{\mu}}$}{o6}
    \fmflabel{${\mu}$}{o3}
    \fmf{fermion}{i1,v1}
    \fmfdot{v1}
    \fmfdot{v1}
    \fmf{fermion}{v1,i2}
    \fmf{boson,label=\begin{scriptsize}${}$\end{scriptsize}}{v1,v5}
    \fmfdot{v1,v5}
    \fmf{fermion,label=\begin{scriptsize}${}$\end{scriptsize}}{v20,v5}
    \fmfdot{v5,v20}
    \fmf{boson,label=\begin{scriptsize}${W}$\end{scriptsize}}{v20,v80}
    \fmfdot{v20,v80}
    \fmfdot{v80}
    \fmf{fermion,tension=0.5}{v80,o3}
    \fmfdot{v80}
    \fmf{fermion,tension=0.5}{o6,v80}
    \fmfdot{v20}
    \fmf{fermion,tension=0.5}{o4,v20}
    \fmfdot{v5}
    \fmf{fermion,tension=0.5}{v5,o5}
  \end{fmfgraph*}}\hfil\allowbreak
\end{fmffile}
\vspace{15pt}
\figcaption{\label{fig:wwtype}The typical topologies of four fermions production in $WW$ type.}
\end{center}

The schedule of classification, which is borrowed from LEP~\cite{ref:17}, for four fermions production depends crucially on the final state. The four fermions can be classified into two classes, the first one comprises the (up, anti-down or $l\bar{\nu}$) and (down anti-up or $\bar{l}\nu$) combinations,
$$(U_i\quad\bar{D}_i)+(\bar{U}_j\quad D_j),$$
which is produced by virtual $WW$ pair, named as ``$WW$'' process. The other class is the production of two fermion-antifermion pairs,
$$(f_i\quad\bar{f}_i)+(\bar{f}_j\quad f_j),$$
which is produced by two virtual neutral vector bosons, named as ``$ZZ$'' process.

A further restriction can be applied to these two types. If there are $e^{\pm}$ together with its neutrino and an on-shell W  boson  in the final state,
this type is named as ``Single $W$'' process;
Meanwhile, if there are a electron-positron pair and a on-shell $Z$ boson  in the final state, this case is named as ``Single $Z$''.
Some final states consists of two mutually charge conjugated fermion pairs, which could be from both virtual $WW$ or $ZZ$, this type is called as ``mixed type''.

The typical structure of Feynman diagrams for WW type is listed in Fig.~\ref{fig:wwtype}, the final states could be produced through intermediate W pair  or a W boson radiation.
Further, the actual number of the Feynman diagrams is listed in Tab.~\ref{tab:ww}. The number in \textbf{bold} font is the general WW process, which means there are two pairs of fermions in final state without identical particles. The ordinary font and \emph{italic} font describe the single $W$ and mixed processes, respectively. The $ZZ$ type has a similar structure with the $WW$ type, and \cite{ref:17} is a good reference for details.

\section{Summary}
\begin{center}
\includegraphics[width=\linewidth]{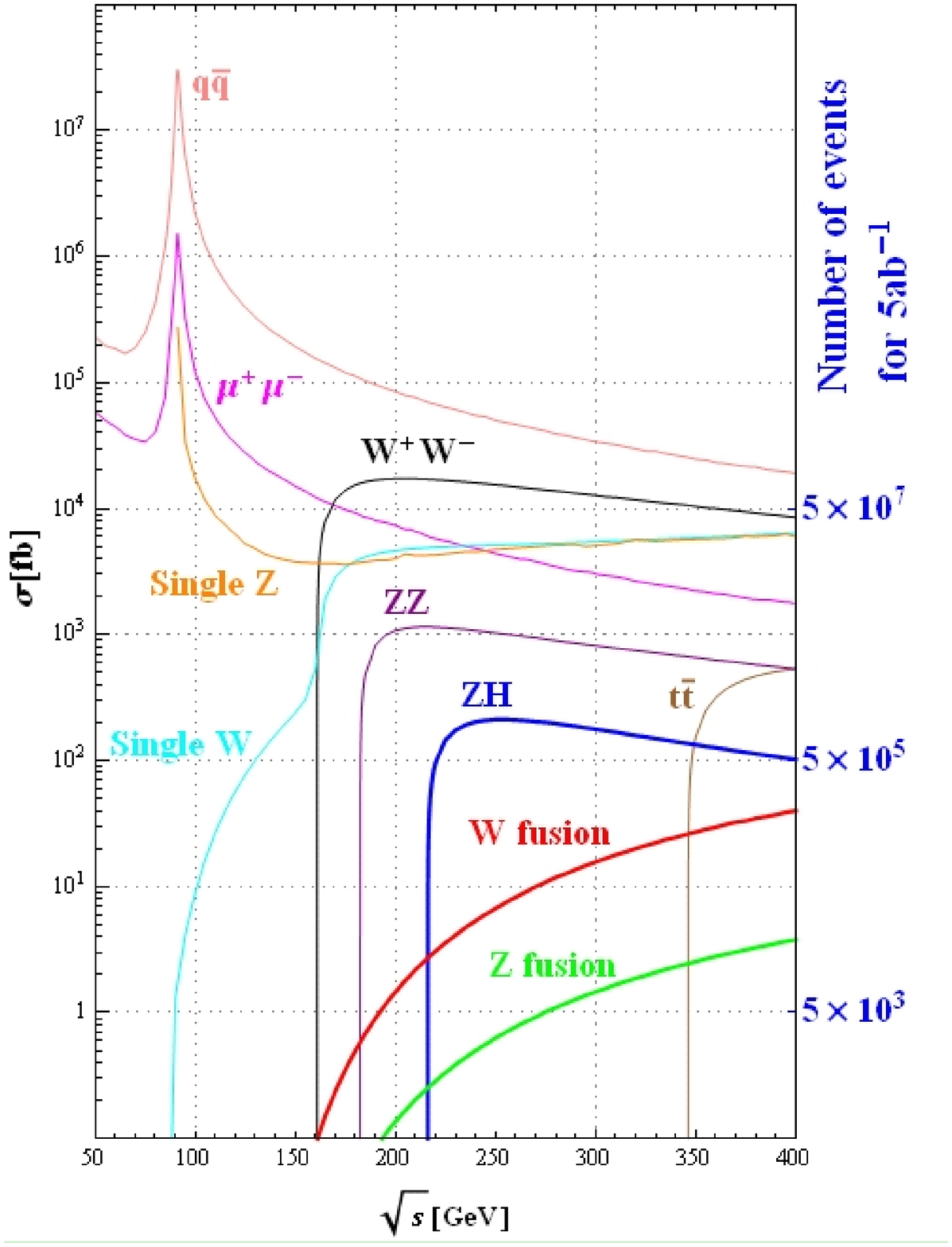}
\figcaption{\label{fig:xs} The cross sections of major processes of SM with ISR effect taken into account.}
\end{center}

In summary, the cross sections of major Standard Model processes, including Higgs production as well as the major backgrounds, are plotted in Fig.~\ref{fig:xs},
where the ISR effect has been taken into account.

In addition the numerical results of these processes are listed in Tab.~\ref{tab:sam},
as well as the expected number of events in total  luminosity of 5 ab$^{-1}$ for a 10-years run.
According to the cross sections, the Monte-Carlo samples for Higgs analysis on CEPC have been generated by {\sc Whizard}.

\begin{center}
\tabcaption{The cross sections and number of events expected at 250 GeV for CEPC\label{tab:sam}}
\begin{tabular}{@{}*{3}{lcc}}
\toprule
Process         & Cross section & No. of events in 5ab$^{-1}$ \\ \hline
\multicolumn{3}{c}{Higgs production cross section in fb } \\ \hline
$e^+e^-\to ZH$ & 212 & $1.06\times10^6$  \\
$e^+e^-\to \nu\bar{\nu}H$ & 6.27 & $3.36\times10^4$ \\
$e^+e^-\to e^+e^-H$ & 0.63 & $3.15\times10^3$ \\ \hline
Total & 219 & $1.10\times10^6$ \\
\\
\multicolumn{3}{c}{Background cross sections in pb } \\ \hline
$e^+e^-\to e^+e^-$ & 25.1 & $1.3\times10^8$ \\
$e^+e^-\to qq$ & 50.2 & $2.5\times10^8$ \\
$e^+e^-\to \mu\mu(\mathrm{or} \tau\tau)$ & 4.40 & $2.2\times10^7$ \\
$e^+e^-\to WW$ & 15.4 & $7.7\times10^7$ \\
$e^+e^-\to ZZ$ & 1.03 & $5.2\times10^6$ \\
$e^+e^-\to eeZ$ & 4.73 & $2.4\times10^7$ \\
$e^+e^-\to e\nu W$ & 5.14 & $2.6\times10^7$ \\
\toprule
\end{tabular}
\end{center}

In this paper, the cross sections  of Higgs production and the background processes at the CEPC are evaluated and the classification for the MC samples is discussed.
Most of the processes are well calculated by {\sc Whizard}. Besides that, Bhabha process should be studied more carefully in the future.

\end{multicols}

\clearpage

\end{document}